\documentclass[twocolumn,tighten]{aastex62}

\usepackage{amsmath}
\usepackage{xspace}
\usepackage{multirow}
\usepackage{fancyapj}
\usepackage{lipsum}
\usepackage{apjfonts}
\usepackage{soul}
\usepackage[normalem]{ulem}

\newcommand{\msol}{\ensuremath{M_{\odot}}\xspace}
\newcommand{\rxte}{\textit{RXTE}\xspace}
\newcommand{\nustar}{\textit{NuSTAR}\xspace}
\newcommand{\nicer}{\textit{NICER}\xspace}

\newcommand{\swift}{\textit{Swift}\xspace}
\newcommand{\source}{Swift~J0243.6+6124\xspace}

\shortauthors{Jaisawal et al.}

\begin{document}

\title{An evolving broad iron line from the first Galactic ultraluminous X-ray pulsar Swift~J0243.6+6124}

\author[0000-0002-6789-2723]{Gaurava K. Jaisawal}\email{gaurava@space.dtu.dk}
\affil{National Space Institute, Technical University of Denmark, 
  Elektrovej 327-328, DK-2800 Lyngby, Denmark}

\author[0000-0002-8585-0084]{Colleen A. Wilson-Hodge}
\affil{ST12 Astrophysics Branch, NASA Marshall Space Flight Center, Huntsville, AL 35812, USA}

\author[0000-0002-9378-4072]{Andrew C. Fabian}
\affil{Institute of Astronomy, University of Cambridge, Madingley Road, Cambridge CB3 0HA, UK}

\author[0000-0003-2865-4666]{Sachindra Naik}
\affiliation{Astronomy and Astrophysics Division, Physical Research Laboratory, Navrangapura, Ahmedabad - 380009, Gujarat, India}

\author[0000-0001-8804-8946]{Deepto Chakrabarty}
\affil{MIT Kavli Institute for Astrophysics and Space Research, 
  Massachusetts Institute of Technology, Cambridge, MA 02139, USA}
  
\author[0000-0001-9840-2048]{Peter Kretschmar}
\affiliation{European Space Astronomy Centre (ESAC), Science Operations Departement, 28692 Villanueva de la Ca\~nada, Madrid, Spain}  

\author[0000-0001-8128-6976]{David R. Ballantyne}
\affiliation{Center for Relativistic Astrophysics, School of Physics, Georgia Institute of Technology, 837 State Street, Atlanta, GA 30332-0430, USA}

\author[0000-0002-8961-939X]{Renee M. Ludlam}
\affil{Department of Astronomy, University of Michigan, 1085 South University Ave, Ann Arbor, MI 48109-1107, USA}
\affil{Cahill Center for Astronomy and Astrophysics, California Institute of Technology, Pasadena, CA 91125, USA}\thanks{Einstein Fellow}

\author[0000-0002-4397-8370]{J{\'e}r{\^o}me Chenevez}
\affil{National Space Institute, Technical University of Denmark, 
  Elektrovej 327-328, DK-2800 Lyngby, Denmark}

\author[0000-0002-3422-0074]{Diego Altamirano}
\affiliation{Physics \& Astronomy, University of Southampton, 
  Southampton, Hampshire SO17 1BJ, UK} 
  
\author{Zaven Arzoumanian} 
\affiliation{Astrophysics Science Division, 
  NASA's Goddard Space Flight Center, Greenbelt, MD 20771, USA}

\author[0000-0003-0388-0560]{Felix F\"{u}rst}
\affiliation{European Space Astronomy Centre (ESAC), Science Operations Departement, 28692 Villanueva de la Ca\~nada, Madrid, Spain}

\author{Keith C. Gendreau} 
\affiliation{Astrophysics Science Division, 
  NASA's Goddard Space Flight Center, Greenbelt, MD 20771, USA}

\author[0000-0002-6449-106X]{Sebastien Guillot} 
\affil{CNRS, IRAP, 9 avenue du Colonel Roche, BP 44346, F-31028 Toulouse Cedex 4, France} 
\affil{Universit\'e de Toulouse, CNES, UPS-OMP, F-31028 Toulouse, France} 

\author[0000-0002-0380-0041]{Christian Malacaria}
\affiliation{NASA Marshall Space Flight Center, NSSTC, 320 Sparkman Drive, Huntsville, AL 35805, USA}\thanks{NASA Postdoctoral Fellow}
\affiliation{Universities Space Research Association, NSSTC, 320 Sparkman Drive, Huntsville, AL 35805, USA}

\author{Jon M. Miller}
\affil{Department of Astronomy, University of Michigan, 1085 South University Ave, Ann Arbor, MI 48109-1107, USA}

\author[0000-0002-5041-3079]{Abigail L. Stevens}
\affiliation{Department of Physics \& Astronomy, Michigan State University, 567 Wilson Road, East Lansing, MI 48824, USA}
\affil{Department of Astronomy, University of Michigan, 1085 South University Ave, Ann Arbor, MI 48109-1107, USA}
\altaffiliation{NSF Astronomy \& Astrophysics Postdoctoral Fellow}

\author[0000-0002-4013-5650]{Michael T. Wolff}
\affil{Space Science Division, U.S. Naval Research Laboratory, Washington, DC 20375, USA}

\begin{abstract}

We present a spectral study of the ultraluminous Be/X-ray transient pulsar Swift~J0243.6+6124 using  {\it Neutron Star Interior Composition Explorer} (\nicer) observations during the system's 2017--2018 giant outburst. The 1.2--10~keV energy spectrum of the source can be approximated with an absorbed cut-off power law model. We detect strong, luminosity-dependent emission lines in the 6--7 keV energy range. A narrow 6.42 keV line, observed in the sub-Eddington regime, is seen to evolve into a broad Fe-line profile in the super-Eddington regime. Other features are  found at 6.67 and 6.97~keV in the Fe-line complex. An asymmetric broad line profile, peaking at 6.67 keV, is possibly due to Doppler effects and gravitational redshift. The 1.2--79~keV broadband spectrum from \nustar and \nicer observations at the outburst peak is well described by an absorbed cut-off power law plus multiple Gaussian lines and a blackbody component. Physical reflection models are also tested to probe the broad iron line feature.  Depending on the mass accretion rate, we found emission sites that are evolving from $\sim$5000~km to a range closer to the surface of the neutron star.  Our findings are discussed in the framework of the accretion disk  and its implication on the magnetic field,  the presence of optically thick accretion curtain  in the magnetosphere, jet emission, and the massive, ultra-fast outflow expected at super-Eddington accretion rates. We do not detect any signatures of a cyclotron absorption line in the {\em NICER} or {\em NuSTAR} data.

\end{abstract}


\section{Introduction}\label{sec:intro}

Fluorescent emission lines of iron have been ubiquitously 
observed at energies in the 6.4--6.9~keV range in the spectra of various 
classes of astrophysical sources, such as X-ray binaries and active galactic 
nuclei (for reviews see \citealt{Fabian2000, Miller2007, 
Bhattacharyya2010}). X-ray ``reflection'' from an accretion disk or stellar 
wind is known to produce narrow Fe emission features. As the line-emitting 
region approaches a compact object, Doppler and relativistic effects 
broaden the Fe line  \citep{Fabian1989}. An asymmetric, red-shifted 
profile emerges in the spectrum due to strong relativistic effects in the vicinity 
of the central X-ray source. Hence, understanding the characteristic properties 
of such broadened emission lines serves to probe the dynamics of the accretion 
flow and the gravitational redshift near the compact object. Asymmetric, 
broad iron profiles are known in active galactic nuclei as well 
as low mass X-ray binaries with black-hole or neutron star accretors 
(see, e.g., \citealt{Miller2006, Miller2007, Bhattacharyya2010}). We study the 
Fe emission in the case of a high-mass X-ray binary pulsar, \source, 
using \nicer observations.

The X-ray transient \source was discovered with the Neil Gehrels 
\swift Observatory in October 2017 during the onset of a strong 
outburst (\citealt{Cenko_2017, Kennea_2017}). Detection of 9.8~s 
pulsations identified the source as an X-ray pulsar (\citealt{Jenke_2017, 
Kennea_2017, Bahramian2017, Jaisawal_2018}). \source harbors a highly magnetized 
neutron star ($\le$10$^{13}$~G; \citealt{Tsygankov2018, Wilson2018}), 
accreting from a massive optical companion of Oe- or early Be-type 
\citep{Bikmaev_2017}. The constraint on its magnetic field is obtained 
tentatively using the independent methods based on the `propeller' luminosity 
from the source flux evolution \citep{Tsygankov2018}, measuring the critical 
luminosity from hardness ratios, and also  from the detected quasi-periodic 
oscillations in \nicer data \citep{Wilson2018}.  The system has a 27.6~d eccentric 
($e\approx \rm 0.1$) orbit \citep{Doroshenko2018, Wilson2018}, and is located at a distance of 
$\approx$7~kpc based on Gaia DR2 as described in \citet{Wilson2018}.

The 2017 giant outburst from the system lasted approximately five 
months, with a peak X-ray luminosity $\sim$10$^{39}$~erg~s$^{-1}$, 
which exceeds the Eddington limit of a neutron star (\citealt{Tsygankov2018, 
Wilson2018}). This luminosity classified the source as an 
ultraluminous X-ray (ULX) pulsar, the very first detected inside our own
Galaxy. ULXs are powerful, off-nuclear (extragalactic), point-like
sources, emitting at a luminosity $\ge$10$^{39}$~erg~s$^{-1}$ 
(see \citealt{Kaaret2017} for a review). A few of them are pulsars 
with spin period in the range $\sim$1 to 30~s  (e.g., \citealt{Bachetti2014, Furst2016, 
Israel2017a}) and magnetic field in the range 10$^{11}$--10$^{13}$~G 
\citep{King2019}.  A much stronger multipolar  field strength is also 
reported in ULX pulsars  (see, e.g. \citealt{Israel2017a, Israel2017b}). 
It remains unclear what powers these sources. A 
possible explanation comes from a combination of super-Eddington 
accretion and geometric collimation effect (\citealt{King2009, 
Middleton2015}). Some of the ULXs are also known to produce relativistic 
jets (\citealt{Kaaret2003} and references therein).

A radio counterpart of \source was detected in the 6 and 22 GHz bands 
using the Karl G. Jansky Very Large Array (VLA; 
\citealt{Eijnden2018}). The analysis suggested that the pulsar 
launches a jet at high mass accretion rate. Current 
understanding favors the jet outflows from black holes (in the
low/hard state) or weakly magnetized neutron stars in X-ray 
binaries (\citealt{Fender2006book}), while the absence
of jet formation in accreting pulsars is thought to be due to 
strong magnetic fields of order $\sim$10$^{12}$~G as well as 
their slower rotation \citep{Massi2008, Parfrey2016}. 
The existence of such strong  field is expected to
hinder the magneto-centrifugal jet driving mechanism by simply  
truncating the accretion disk far away from the neutron star.

The observed radio emission from \source evolved across the 2017 
outburst. It followed a radio-X-ray correlation, 
$L_{\rm radio}\propto L_{\rm x}^{0.54\pm0.16}$, 
consistent with both black hole and neutron star X-ray binaries. 
Further VLA monitoring found that the jet emission turned on and off 
within days \citep{Eijnden2019}.  The jet emission also appeared 
when the pulsar re-brightened in fainter X-ray outbursts in 
March and May 2018. The jet persisted in these later observations 
only when the X-ray luminosity was $\ge$4$\times$10$^{36}$ 
($D$/7 kpc)$^2$ erg~s$^{-1}$, and was  equally radio-bright 
as at the peak of the 2017 outburst \citep{Eijnden2019}.  

Using \nicer (\citealt{Gendreau2012}), we explore 
the spectral characteristics of the pulsar  \source\ during the 
2017--2018 strong outburst. This paper mainly focuses on the detection 
of a broad, asymmetric iron emission line and its evolution across 
the sub- and super-Eddington regimes. We describe the observations 
and analysis in Section~2. The results and discussion are presented 
in Section~3 and Section~4, respectively.


\begin{figure}
\centering
\includegraphics[height=2.38in, width=3.35in, angle=0]{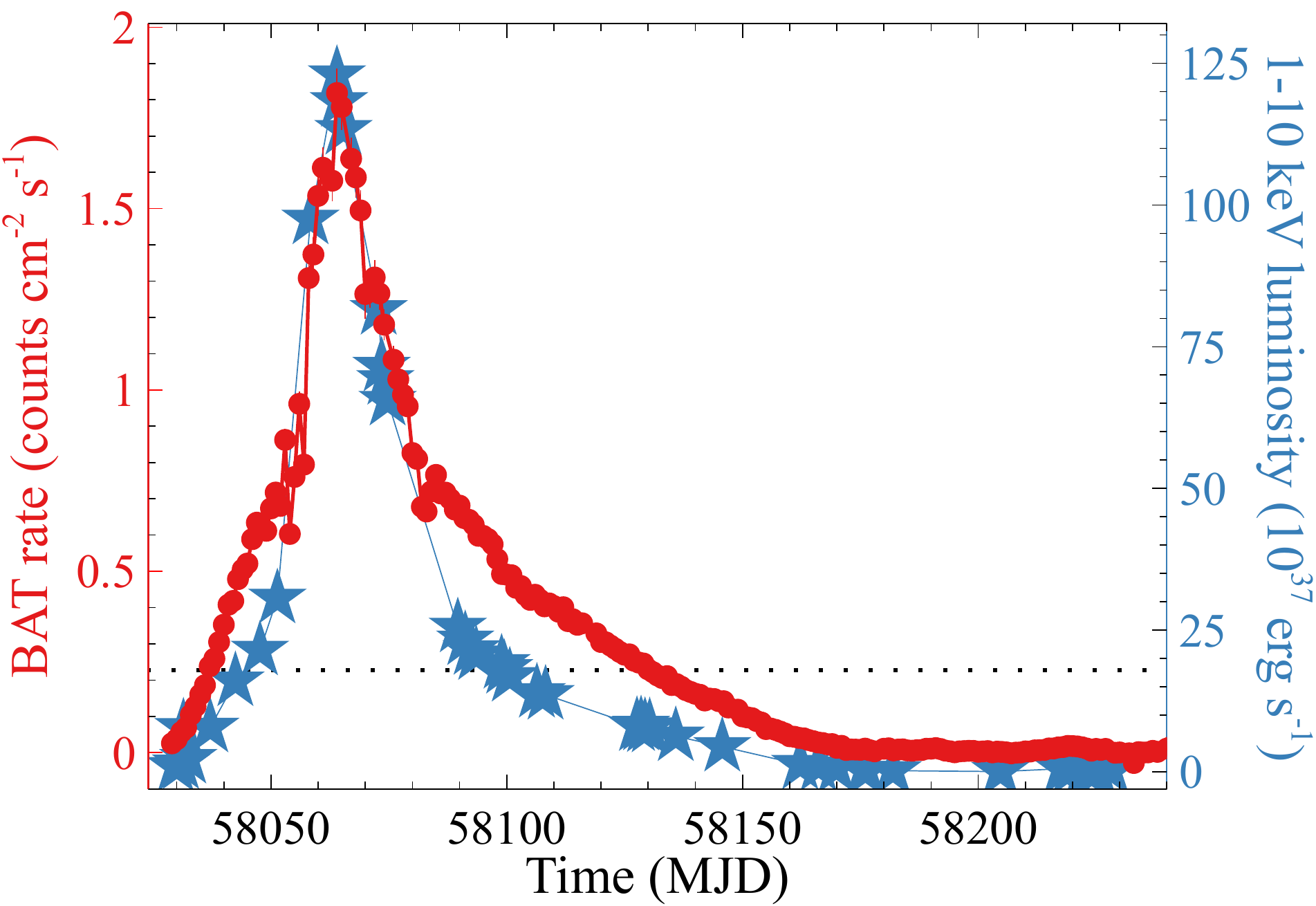}
\caption{Outburst monitoring light curve of \source from \swift/BAT
in the 15-50 keV range (solid dots;  \citealt{Krimm2013}).  
The 1--10 keV unabsorbed luminosity light curve (solid stars) is 
also shown from \nicer observations between 2017 October 03 
(MJD 58029) and 2018 April 20 (MJD 58228). The pulsar was extremely bright at 
the outburst peak, reaching 9 Crab intensity in the BAT band; for reference, 1~Crab represents
0.220 counts~cm$^{-2}$~s$^{-1}$ in this band.  The horizontal dotted-line in the figure 
represents the Eddington luminosity of a 1.4 \msol neutron star}. \label{fig-lc}
\end{figure}

\section{Observations and Data Analysis} \label{sec:obs}

\subsection{\nicer} \label{sec:nicer}

The \nicer X-ray timing instrument (XTI, \citealt{Gendreau2012, 
Gendreau2016}) is a non-imaging, soft X-ray telescope installed 
on the International Space Station in June 2017. The XTI consists of an 
array of 56 co-aligned concentrator optics, each associated with a 
silicon drift detector \citep{Prigozhin2012}, together operating in the 
0.2--12 keV band. The XTI provides high time resolution 
of $\sim$100~ns (rms) and spectral resolution of $\approx$85~eV 
at 1~keV. It has a field of view of $\approx$30~arcmin$^2$ in 
the sky. The effective area of \nicer is $\approx$1900~cm$^2$ 
at $1.5$~keV, with 52 active detectors.  

\nicer has continued monitoring  \source{} since its discovery in 
2017 October by covering numerous epochs of the outburst (Figure~\ref{fig-lc}). 
Initial results from these observations were reported  in \citet{Wilson2018}. 
Here, we present evidence of an evolving iron line using 
the same data sets (Observation IDs 1050390101--1050390160)
between 2017 October and 2018 April. A detailed observation log can be 
found in Table~1 of  \cite{Wilson2018}. 

Data were reduced using \textsc{HEAsoft} version 6.24, 
\textsc{nicerdas} version 2018-04-24\_{V004}, and the 
calibration database (CalDB) version 20180711. We excluded 
events from the South Atlantic Anomaly region; other 
filtering criteria such as elevation angle $>$30$^{\circ}$ 
from the Earth limb, pointing offset $\le$54 arcsec, offset from  
the bright Earth $>40^{\circ}$, and also $\textrm{COR\_SAX}\ge$4 
to remove the high charged particle background regions were  
used to screen the data. Here, COR$\_$SAX is the magnetic 
cut-off rigidity, in GeV/c. From the resulting good time intervals, 
final spectral products were created with \textsc{XSELECT}. 
We used blank-sky regions (e.g., field \rxte-6; \citealt{Jahoda2006})
to estimate the background corresponding to our observations.

\subsection{\nustar} \label{sec:nustar}

Following the source discovery, the {\it Nuclear Spectroscopic Telescope Array} 
(\nustar; \citealt{Harrison2013}) monitored the pulsar at various epochs during 
the 2017--2018 outburst. We note that these observations have already been used 
for spectral studies \citep{Jaisawal_2018, Tao2019}. In this paper, we focus on 
quasi-simultaneous \nustar and \nicer observations from the outburst peak in order 
to study the broad-band X-ray spectrum. The effective exposure of the \nustar pointing  
on 2017 October 31 (ObsID 90302319004) is 1293~s, while the \nicer observation took 
place on 2017 November 1 for an effective exposure of 3406~s (ObsID 1050390113).

Standard methods were adopted for \nustar data 
analysis\footnote{\url{https://heasarc.gsfc.nasa.gov/docs/nustar/analysis/nustar\_swguide.pdf}}. 
Before reprocessing, we set an additional keyword `statusexpr' 
to ``STATUS==b0000xx00xx0xx000'' in \texttt{nupipeline} 
in order to retain events that might have been flagged as ``hot'' or 
``flickering'' due to the high count rate of the source $>$100 counts~s$^{-1}$. 
Data from both focal plane modules, FPMA and FPMB, were used 
in our analysis. We considered a source region of 150~arcsec 
for the spectral extraction, while the background is measured 
from a source-free region with a similar radius. The final products 
were generated using \texttt{nuproduct} task.

\begin{figure}
\centering
\includegraphics[height=3.2in, width=5.3in, angle=-90]{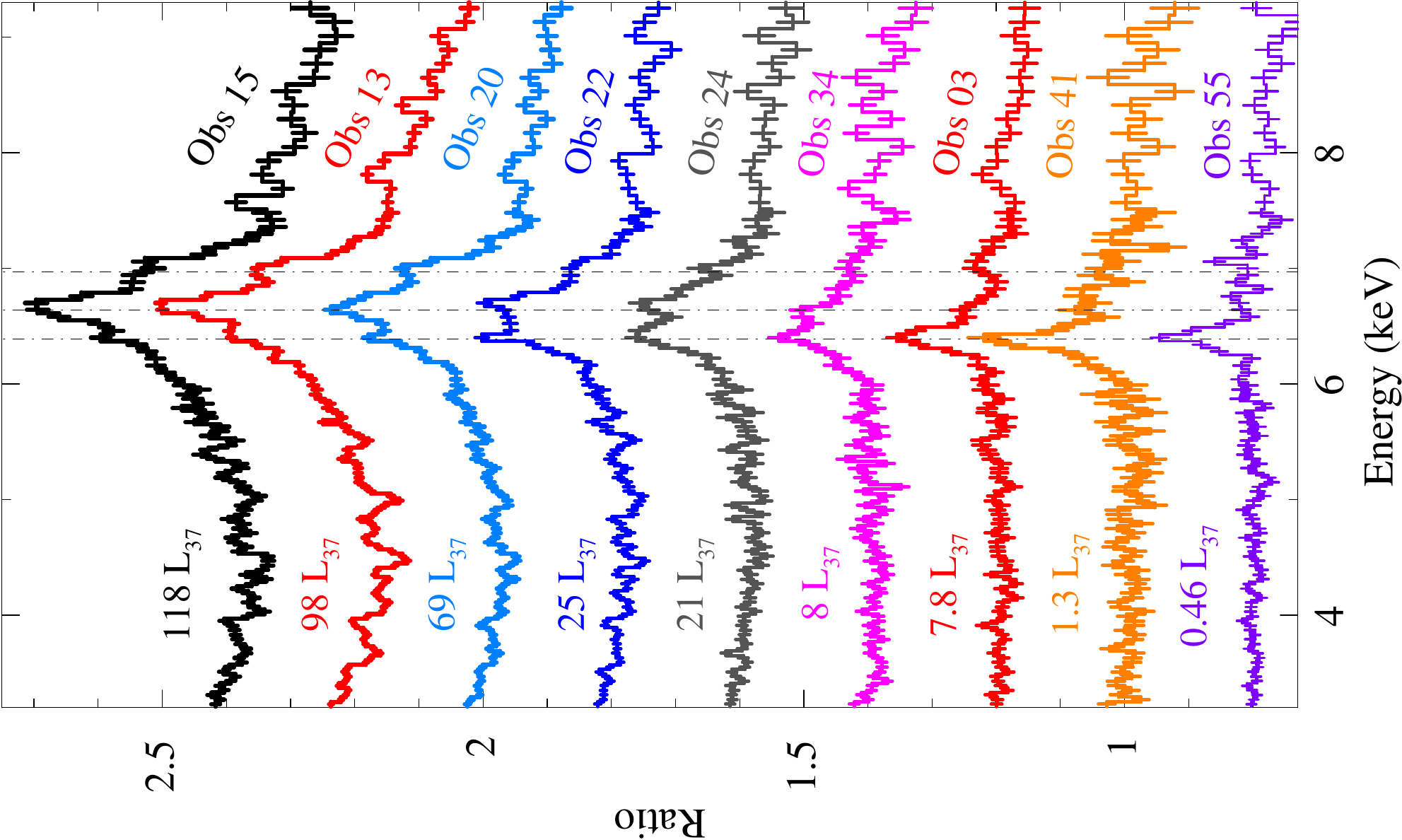}
\caption{Ratio of \nicer data to the continuum model (scaled arbitrarily) obtained 
by fitting 1.2--10 keV energy spectra of \source with an absorbed cutoff power-law model. 
These are arranged in order of increasing luminosity. L$_{\rm 37}$ stands for 
unabsorbed luminosity in units of 10$^{37}$~erg~s$^{-1}$. Observation IDs (10503901xx) 
are also shown as labels. Three emission lines in the 6--7 keV range are detected 
at 6.4, 6.67, and 6.98~keV energies (vertical dash-dot lines).} 
\label{fe-ratio}
\end{figure}

 \begin{figure*}
 \begin{center}$
 \begin{array}{cccc}
 \includegraphics[height=5.85 cm, width=5.25 cm, angle=-90]{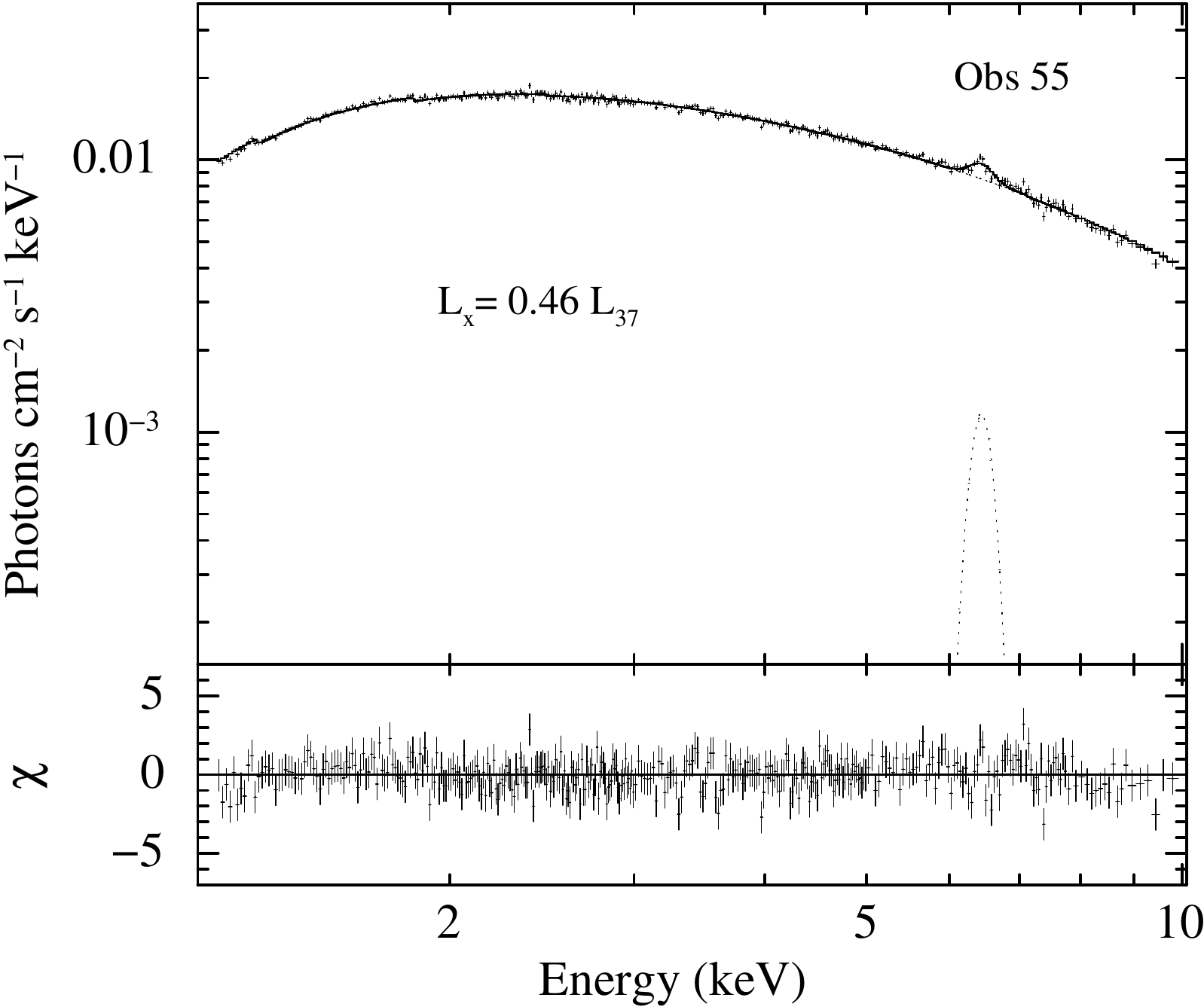} &
 \includegraphics[height=5.85 cm, width=5.25 cm, angle=-90]{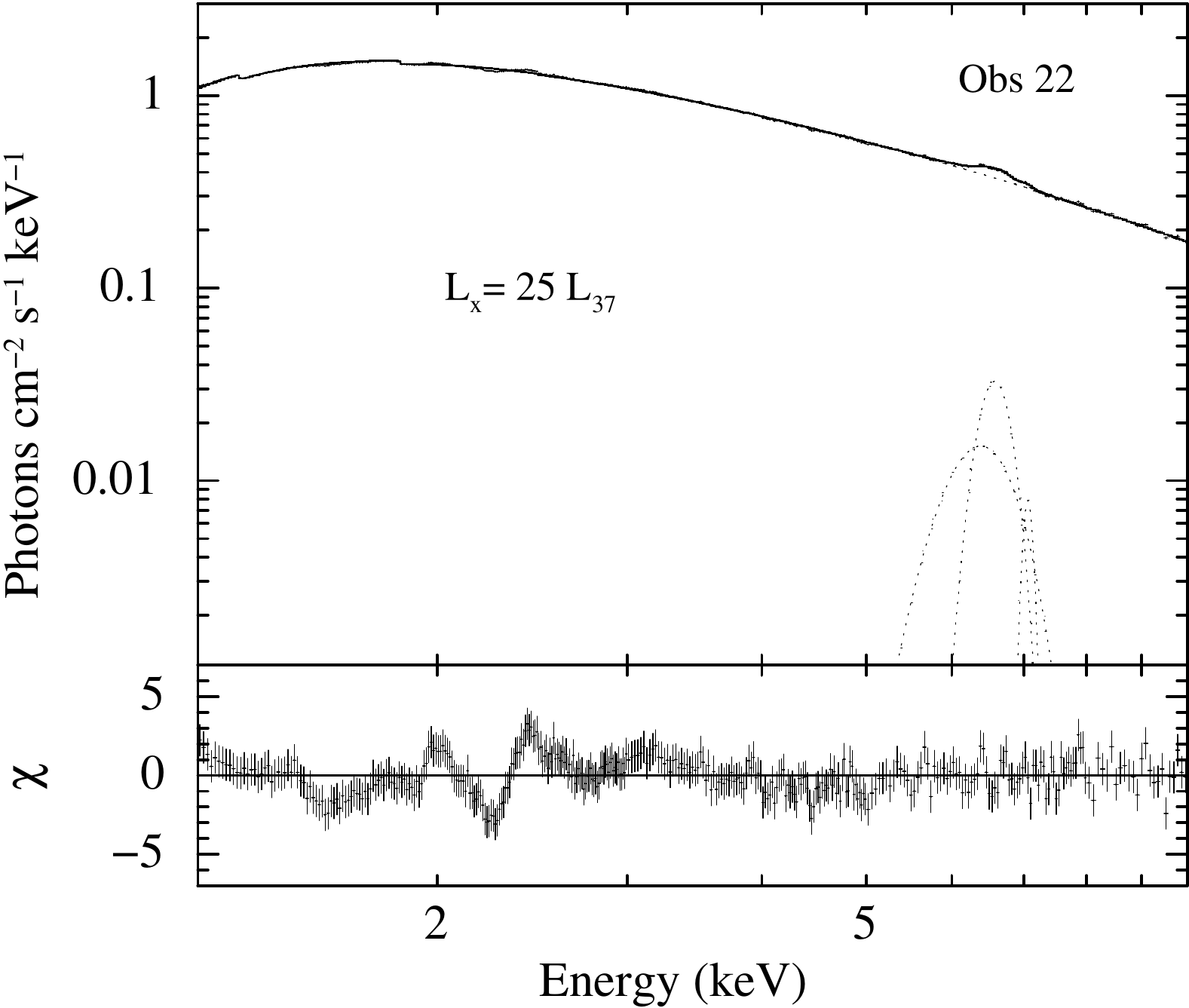} & 
 \includegraphics[height=5.85 cm, width=5.25 cm, angle=-90]{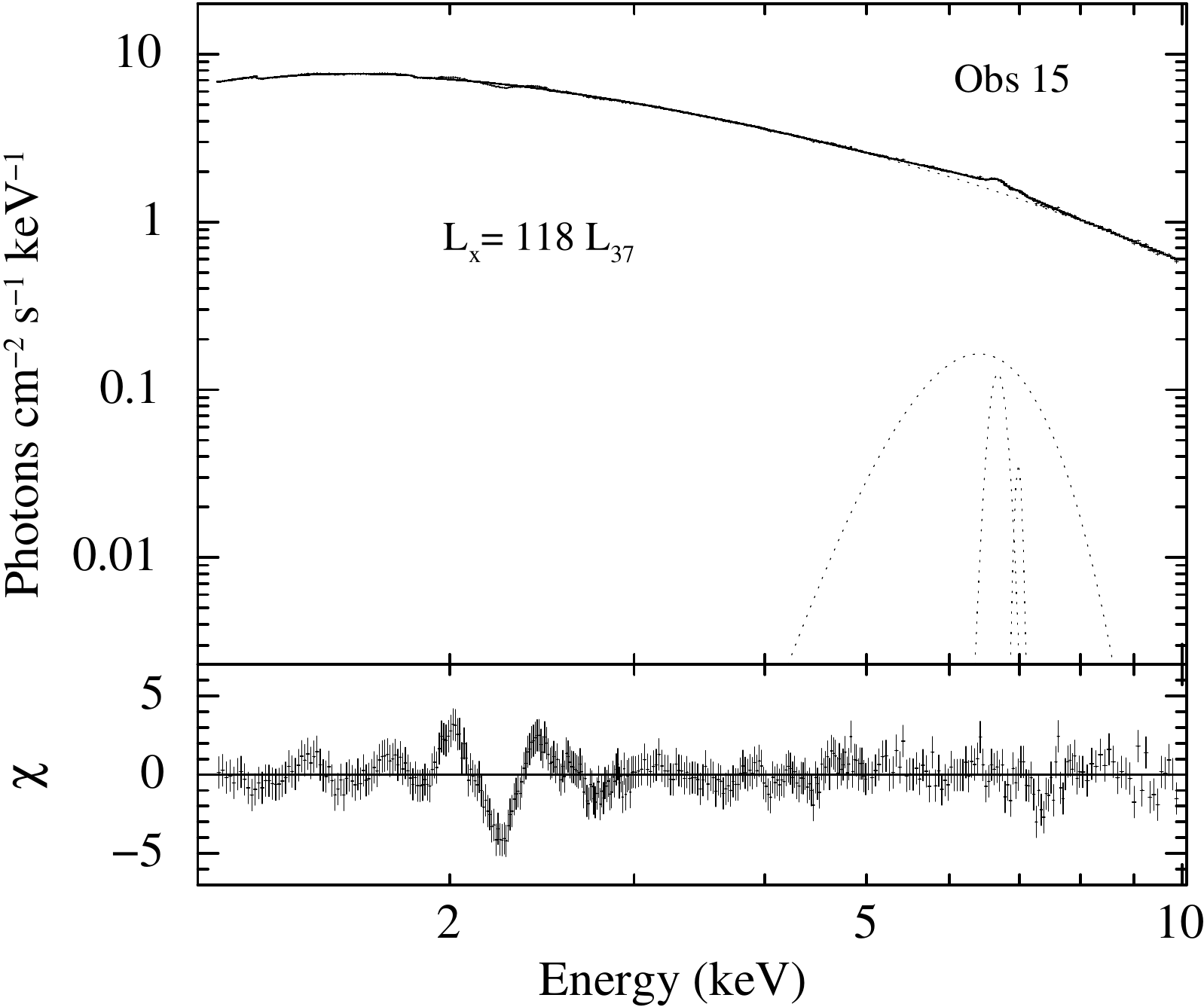} & \\
 \end{array}$
 \end{center}
\caption{The 1.2--10 keV energy spectrum of \source with \nicer at three 
representative luminosities after fitting with an absorbed cutoff power-law 
model plus Gaussian components (top panel). A single narrow line at $\approx$6.4~keV 
was required at the lower luminosity (left; Obs 55). Two emission lines were clearly 
detected in the Fe-band at higher luminosity (middle; Obs 22). The energy spectrum at 
the super-Eddington state is described using an absorbed cutoff power-law model with 
three line components (right; Obs 15). Spectral residuals corresponding to the best-fitting 
models are shown in the bottom panels. The feature observed in the 2--3 keV range of the residuals
(middle and right panels) is possibly instrumental in origin.}
\label{fig2-spec}
\end{figure*}   

\section{Spectral Analysis and Results}

The spectrum was analyzed using \textsc{XSPEC} version 12.10.0 
\citep{Arnaud1996} along with response matrices and effective 
area files (version 1.02 for \nicer). The \nicer spectrum suffers 
from instrumental residuals in the Si (1.7--2.1 keV) and Au (2.2--2.3~keV) 
bands for bright sources like \source. These features are prominent due 
to the high sensitivity of \nicer and calibration uncertainties. To minimize 
the effects of the residuals, we adopted a Crab correction technique that 
exploits the featureless power-law spectrum observed from the Crab Nebula. 
In this approach, the target spectrum is renormalized by a residual 
template created by dividing the observed Crab Nebula spectrum by 
a simulated Crab spectrum (see, e.g., \citealt{Ludlam2018} and 
\citealt{Wilson2018} for details).  The Crab-corrected spectrum of \source 
is fitted after adding a 1.5\% systematic error in the 
1.2--10 keV range.  The data below 1.2 keV are ignored because
of calibration uncertainties and strong residuals at 1 keV.

\subsection{Evolution of iron lines}\label{sec:fe-evo}

To study the evolution of the Fe line complex, we fitted 
each of the \nicer observations with an absorbed
cutoff power-law model. This empirical model is commonly 
employed to describe the continuum spectrum of accretion-powered 
X-ray pulsars \citep{White1983ApJ, Bildsten1997}. A similar model 
with an additional blackbody component has successfully reproduced 
the 3--79~keV \nustar spectrum of \source at a low luminosity 
\citep{Jaisawal_2018}. Considering the limited energy band up to 10~keV, 
the cutoff power-law model, using the {\tt tbabs} absorption 
model with the abundance table of \citet{Wilms2000},   
is preferred to fit the \nicer spectrum.

Figure~\ref{fe-ratio} shows the spectral ratio 
(between the source spectrum and corresponding best-fit continuum model) 
from some of the representative \nicer observations at several epochs 
of the X-ray outburst. The Fe line structure is clearly visible in the 
residuals. A narrow emission line at $\approx$6.4 keV was detected in 
the spectrum at lower source luminosity (e.g., ObsID 55). As the intensity 
increased, this feature became broader in energy, and an additional emission 
component at $\approx$6.97~keV appeared. An emission line at $\approx$6.67~keV 
also becomes apparent at higher luminosity (see, e.g., ObsID 24). 
At even higher luminosity, the line profile transformed into a broad, 
double-peaked shape with almost equal contribution from the 6.4 and 6.67~keV 
lines at a source luminosity of $\approx$25$\times$10$^{37}$~erg~s$^{-1}$ 
(Figure~\ref{fe-ratio}). Finally, the emission from 6.67 keV 
dominated the line complex at a luminosity $\ge$69$\times$10$^{37}$~erg~s$^{-1}$.  
A broad line profile with a strong red wing was also observed in the 4--8~keV range 
in the super-Eddington regime. This is the first time that such a broad asymmetric Fe line 
has been detected in a strongly-magnetized accreting neutron star (see also  
\citealt{Miller2018} and \citealt{Kara2019}  for \nicer spectral
capabilities  in the Fe-band). 
The 6.4 keV line is identified as a fluorescent line from neutral or
weakly ionized Fe, whereas the 6.67 and 6.98 keV lines originate 
from highly ionized He-like and H-like Fe ions, respectively.

We modeled the Fe line structure by adding  Gaussian line components 
to our absorbed cut-off power-law  model. For low source luminosities, 
Gaussian components were added in the 6--7 keV range to fit the observed 
narrow lines (e.g., left and middle panels of Figure~\ref{fig2-spec}). 
 By contrast, a significantly broad line from the ultraluminous phase was 
modeled using three Gaussian functions, keeping their central energies 
fixed at 6.4, 6.7, and 7.0~keV (e.g., right panel of Figure~\ref{fig2-spec}). 
The inclusion of the  emission components in the spectral model improved the 
overall fit with a reduced $\chi^2$ $\le$1.5 in each case  and provided a 
statistical improvement of $>$6~$\sigma$  in the fitting. The statistical 
significance of the emission line is determined by using the {\it XSPEC} 
script {\tt simftest} that estimate the F-test probability through 
Monte Carlo simulations (see, also \citealt{Protassov2002}).  

\begin{figure}
\centering
\includegraphics[height=3.35in, width=2.8in, angle=-90]{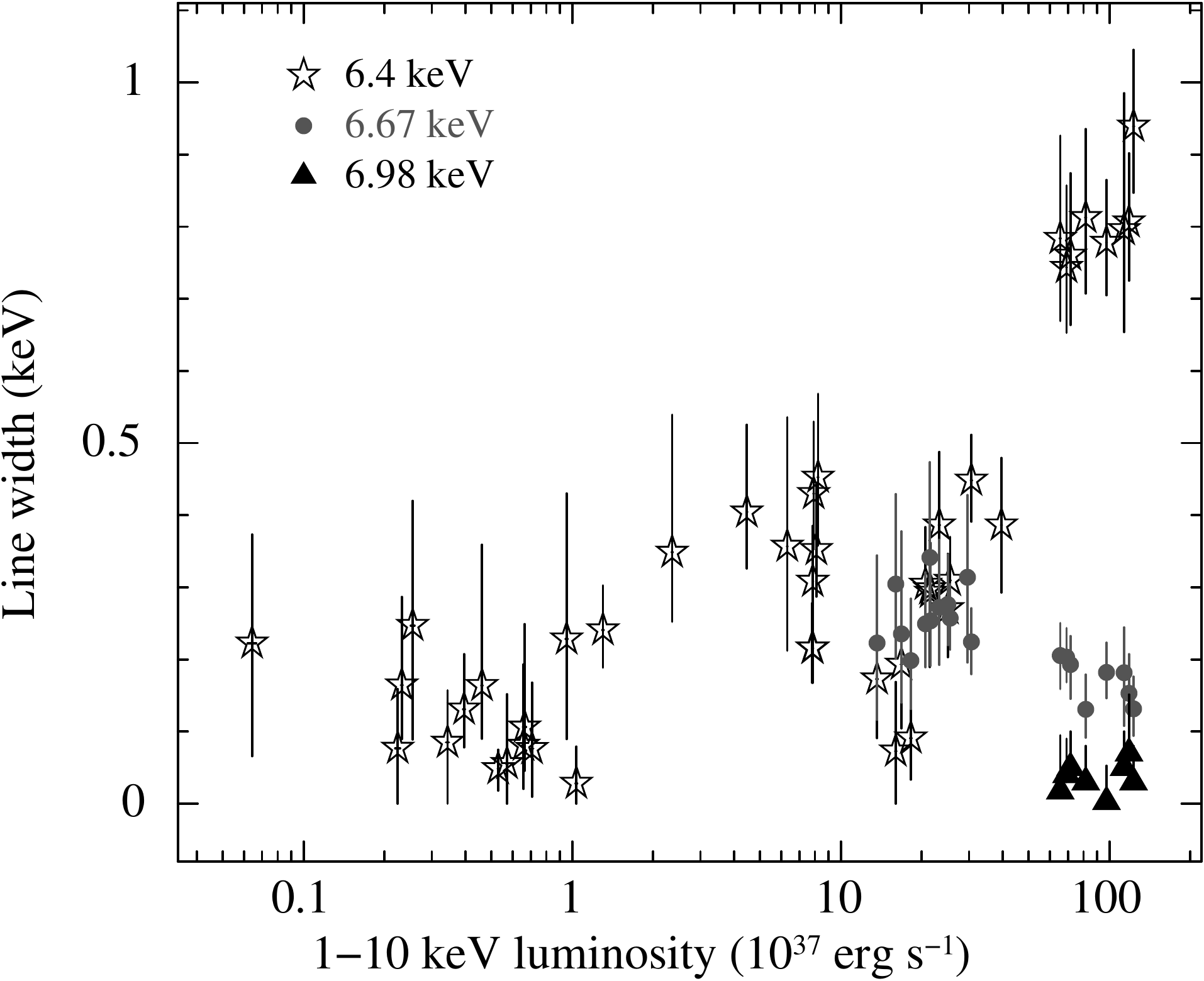}
\caption{ The luminosity evolution of line width for 6.4, 6.67, and 6.98~keV emission 
components during the outburst.} 
\label{line-width}
\end{figure}

\begin{figure}
\centering
\includegraphics[height=3.2in, width=2.8in, angle=-90]{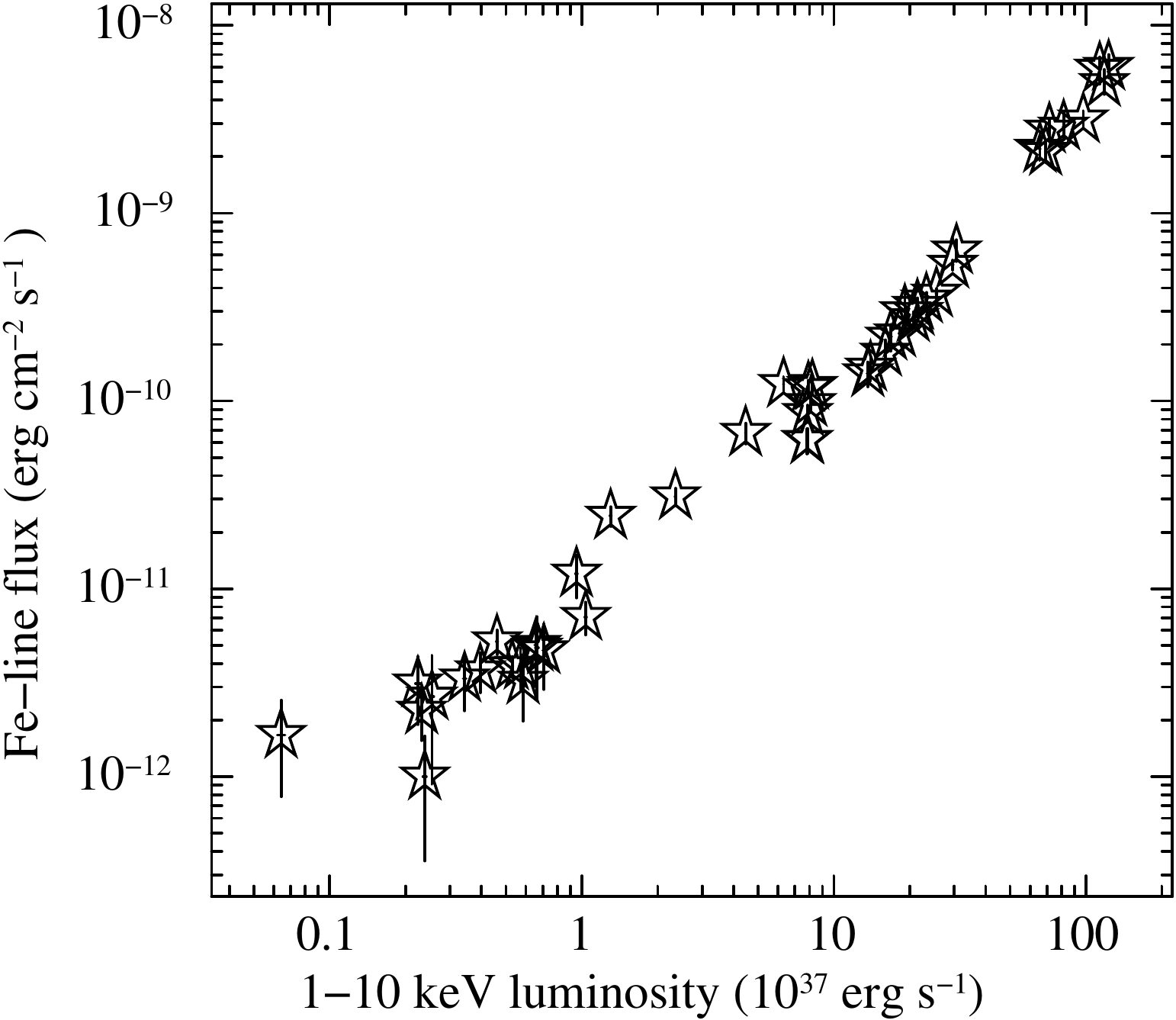}
\caption{Variation in the 1--10 keV unabsorbed luminosity with Fe-band 
flux obtained from fitting \nicer data with a cutoff power-law model. 
The line flux traces the source intensity for more than three orders of magnitude. }
\label{line-flux}
\end{figure}

We employed the {\tt cflux} model to compute the 1--10 keV unabsorbed 
flux. The Fe line fluxes are calculated separately, combining 
all individual Gaussian components in the model spectra. 
The estimated fluxes are also corrected 
for detector deadtime (see Section 3.1 of \citealt{Wilson2018} for 
a discussion on \nicer deadtime corrections for bright sources).
The 1--10 keV source luminosity is calculated assuming a distance of 7 kpc. 
Figure~\ref{line-width} shows the variation between luminosity and the line width 
of each of the Fe-components. A clear evolution from a narrow to broad emission line  
is detected for the 6.4 keV feature in sub- and super-Eddington domains, respectively. 
However, the  6.67 and 6.98~keV lines are remain narrow above the Eddington 
luminosity. The correlation between the source luminosity and the observed 
Fe-line flux is also shown in Figure~\ref{line-flux}.
We found that the Fe-line flux strongly traces the source 
intensity over more than three orders of magnitude. 
We focus on the results from the iron emission line in this paper. A detailed 
discussion of all other parameters will follow in a future publication.

\begin{figure}
\centering
\includegraphics[height=3.in, width=2.8in, angle=-90]{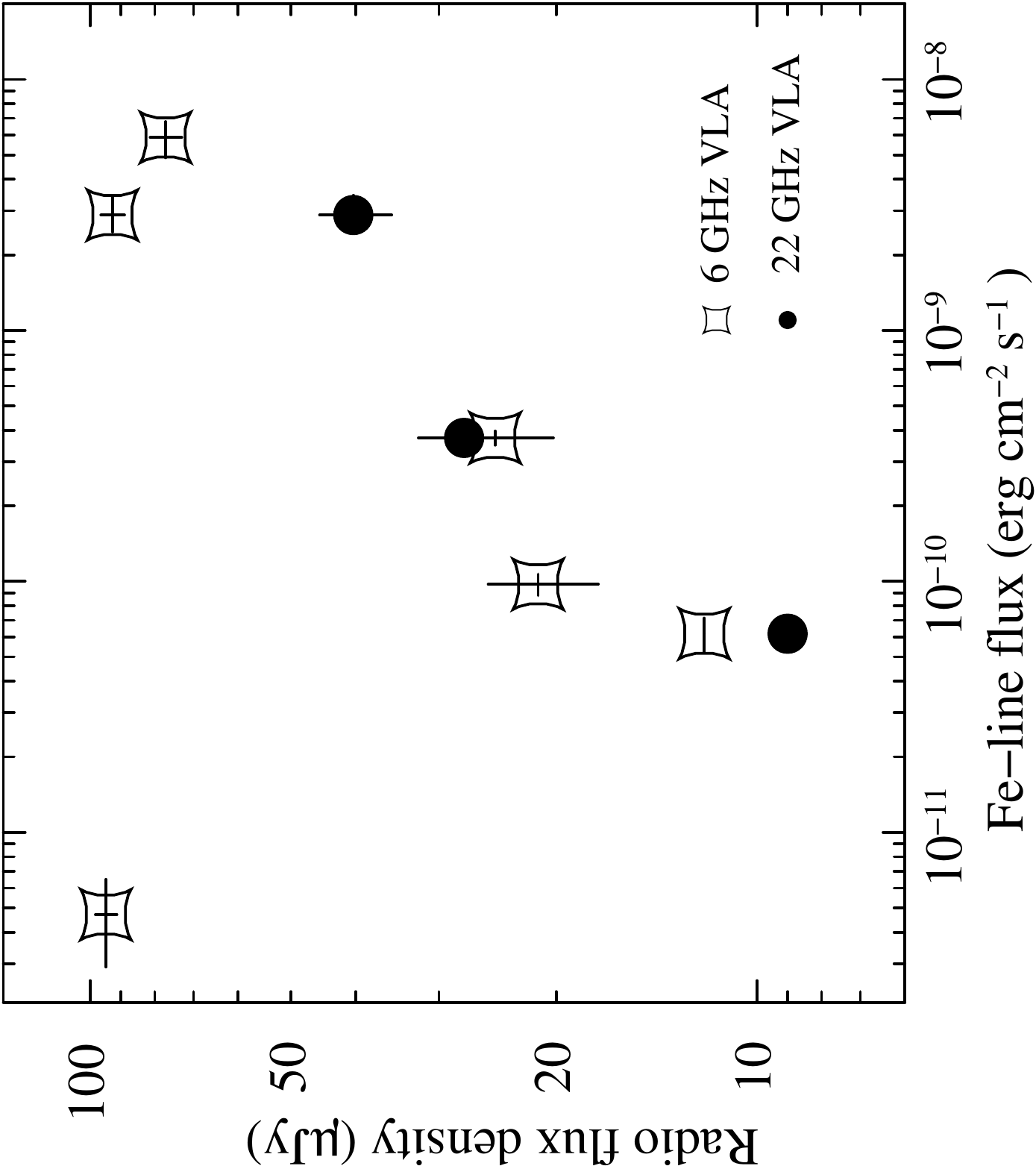}
\caption{Observed iron-band flux with radio flux densities at two frequencies.}
\label{radio-line}
\end{figure}

We also explored the correlation between Fe-line emission and the radio 
flux density as observed with VLA (Figure~\ref{radio-line}). 
The radio data used here are taken from \citet{Eijnden2018} and 
\citet{Eijnden2019}. The VLA observations were quasi-simultaneous 
to \nicer pointings (ObsID 1050390105, 1050390116, 1050390118, 
1050390122, 1050390134, and 1050390143)  between 2017 October 
and 2018 February. From Figure~\ref{radio-line}, it is clear that 
the Fe-line flux shows a positive correlation with radio flux 
density in the high-flux regime. This pattern is possibly changed at 
lowest Fe-line flux (one data point at 6 GHz). The observed variation is 
consistent with the radio-X-ray luminosity relation, in general, reported 
by \citet{Eijnden2018, Eijnden2019}, where the source followed a positive 
trend with radio measurement at high mass accretion rate. A change in 
the correlation was observed at lower luminosity due to the detection of 
jet emission similar in brightness to that at the outburst peak (\citealt{Eijnden2018, Eijnden2019}).

\subsection{Broadband spectroscopy with \nicer and \nustar}

\begin{figure}
\centering
\includegraphics[height=3.3in, width=4.2in, angle=-90]{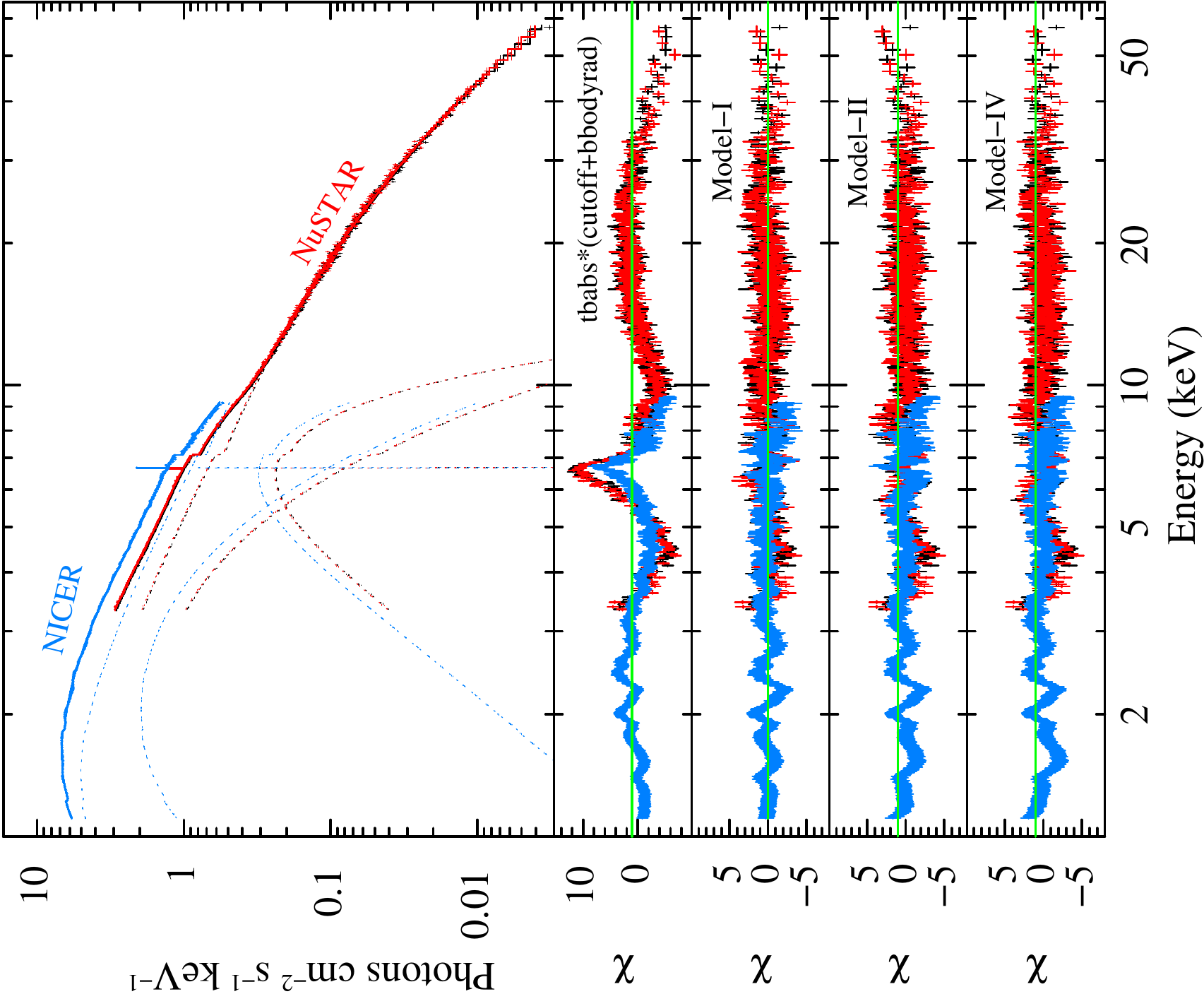}
\caption{The 1.2--79 keV energy spectrum of \source obtained from 
quasi-simultaneous \nicer and \nustar observations at the peak of the 2017 outburst along with (top and middle panels)
the best-fit model comprising a cutoff power-law continuum with a
blackbody component, three Gaussian functions, and an edge feature (Model-I). 
Spectral residuals corresponding to additional models,  \textsc{tbabs*(cutoff+bbodyrad)},  
Model-II: \textsc{tbabs*(gsmooth*Coplrefl+hecut+bbodyrad)}, 
and Model-IV: \textsc{tbabs*(relxill+cutoff+bbodyrad)} are shown in the remaining 
panels. The cross normalization constant between \nicer and \nustar is 1.3 for 
the brightest observation of the source. The value is obtained with respect to the
\nustar/FPMA detector.}   
\label{spec}
\end{figure}

We used data from quasi-simultaneous \nicer and \nustar observations 
at an epoch close to the outburst peak to understand the broadband emission
characteristics of \source in the super-Eddington domain. The 1.2--79 keV 
energy spectrum was first fitted with an absorbed cutoff power-law (Cutoff) with 
a blackbody (BB) component for excess emission in the soft X-rays \citep{Jaisawal_2018}. 
As expected,  the above model yielded a statistically unacceptable fit due to a strong, 
asymmetric, broad emission line complex in the 4--8 keV range (second panel of Figure~\ref{spec}). 
The goodness of the fit  is $\chi^2_{\nu}$ = $\chi^2$/$\nu$ = 5.33, where $\nu$
indicates the number of degrees of freedom (dof). This also produced strong residuals 
or spectral curvatures at both ends of the  4--8~keV band (see second panel of Figure~\ref{spec}). 
Recently, \citet{Tao2019}  approximated the same \nustar data in 3--79~keV 
with a complex model. The authors used a cutoff power-law continuum along 
with three blackbody components, at 0.57, 1.46, and 4.5~keV temperatures. The thermal blackbody
at 0.57 keV (with a radius of 120~km) is interpreted as the photospheric 
emission from the optically-thick outflow, while the components at 1.46~keV  
with radius 20~km and 4.5~keV with radius 1.4~km are explained as a 
temperature gradient from the top to the base of the accretion column, 
respectively \citep{Tao2019}. A marginal detection of an asymmetric Fe-line 
was also reported in the above work.

Combining the \nustar data with the high sensitivity and soft 
X-ray coverage of \nicer, the 1.2--79 keV energy spectrum 
was fitted using a Cutoff+BB model along with three Gaussian 
functions  of fixed central energies at 6.4,  6.7, and 
7.0~keV (see Table~\ref{spec-para}). An edge feature at 
$\approx$7.1~keV is also added in the  model. This provides 
a goodness of fit $\chi^2$/$\nu$ = 1.14 for 2556 dof
(third panel of Figure~\ref{spec}; Model-I). It is important to 
note that the addition of the Gaussian components within the Fe band resolved 
the hump-like residuals that appeared in the Cutoff+BB model fit  with a statistical 
significance of $\approx$100~$\sigma$ in the $\chi^2$ value. 
Interestingly, \citet{Tao2019}  fitted these wavy features using 
thermal blackbodies with temperatures of 1.46 and 4.5~keV. In their study, the peaked 
emissions (3~$kT_{\rm bb}$) of these thermal components accommodated 
the Fe-band features that we have identified as lines  
(see Figure~2 of \citealt{Tao2019}).

\begin{singlespace}
\begin{table*}
\centering
\caption{Best-fit spectral parameters obtained from fitting 
 \nicer and \nustar data close to the outburst peak with 
Model-I: \textsc{tbabs*(cutoff+bbodyrad+ga1+ga2+ga3)*edge}, 
Model-II: \textsc{tbabs*(gsmooth*Coplrefl+hecut+bbodyrad)}, 
Model-III: \textsc{tbabs*(gsmooth*xillver+cutoff+bbodyrad)}, 
and Model-IV: \textsc{tbabs*(relxill+cutoff+bbodyrad)}.}
\small{
\begin{tabular}{llccccc}
\hline
Model       &Parameter          &Model-I            &Model-II        &Model-III     &Model-IV     \\
\hline
\textsc{tbabs} &N$_{\rm H}$$^a$         &0.86$\pm$0.01      &1.67$\pm$0.08    &1.23$\pm$0.02     &0.84$\pm$0.01\\
\textsc{bbodyrad}    &kT (keV)          &0.85$\pm$0.01      &1.16$\pm$0.04    &1.12$\pm$0.02   &1.12$\pm$0.01\\
               &norm                    &4337$\pm$30        &886$\pm$150          &843$\pm$32      &818$\pm$28\\
\textsc{Power-law} &$\Gamma$            &1.35$\pm$0.01       &1.88$\pm$0.03     &1.61$\pm$0.02  &1.65$\pm$0.01\\
\textsc{cutoff} &E$_{\rm cut}$ (keV)    &21.8$\pm$0.3       &26$\pm$2      &27.3$\pm$0.7 &27.2$\pm$0.7\\
                 &E$_{\rm fold}$ (keV)  &...                &8.2$\pm$0.1      &... &...\\

\textsc{Gaussian1}  &E$_c$ (keV)              &6.42               &...  &...   &...\\
                &$\sigma$ (keV)         &1.67$\pm$0.01      &...  &...  &...\\
                &norm                    &0.99$\pm$0.01      &...   &...   &...\\      
\textsc{Gaussian2}  &E$_c$ (keV)              &6.67     &...   &...  &...\\
                &$\sigma$ (keV)         &0.05$\pm$0.02     &...   &...  &...\\
                &norm                   &0.02$\pm$0.01       &...     &...   &...\\      
\textsc{Gaussian3}  &E$_c$ (keV)              &6.98               &...  &...   &...\\
                &$\sigma$ (keV)         &0.1                &...   &...  &...\\
                &norm (10$^{-4}$)       &2$\pm$1            &...    &...   &...\\      
\textsc{Edge}  &Energy (keV)                  &7.11$\pm$0.01      &...   &... &... \\
               &$\tau$                  &0.12$\pm$0.01      &...    &... &... \\

\textsc{Gsmooth}  &$\sigma_{\rm 6 keV}$ (keV)    &...    &0.33$\pm$0.02  &0.19$\pm$0.03     &...        \\

\textsc{Coplrefl}  &$\Gamma$            &...    &0.87$\pm$0.07  &...     &...        \\
          &log($\xi$)                   &...    &2.61$\pm$0.08  &...    &...   \\ 
            &$z$                     &...       &0.0375$\pm$0.002  &... &...     \\
        & norm (10$^{-28}$)             &...    &2.3$\pm$0.6    &...    &...    \\

\textsc{Xillver}   &$\Gamma$            &...     &... &1.18$\pm$0.05       &...        \\
          &log($\xi$)                   &...     &...  &3.47$\pm$0.04      &...   \\ 
          & $A_{Fe}$                    &...     &... &5.1$\pm$0.3      &...\\
               &$z$                     &... &...      &0.0219$\pm$0.003 &...     \\
                &$i(^\circ)$                 &... &...     &23$^{+28}_{-23}$ &... \\
        & norm (10$^{-2}$)              &...     &... &5.0$\pm$0.5     &...    \\

\textsc{Relxill}          &log($\xi$)                   &...    &...   &...   &3.49$\pm$0.03   \\ 
         &$A_{Fe}$                       &...  &...   &...    &4.9$\pm$0.3\\
               &$z$                     &... &... &...    &0.0224$\pm$0.007 \\
           &$i(^\circ)$                 &... &... &...    &43$^{+25}_{-10}$ \\
            &$R_\textrm{in}$ ($R_{g}$)         &... &... &...   &253$^{+237}_{-121}$ \\
           &$f_\textrm{refl}$            &... &... &...   &4.0$^{+0.2}_{-8}$ \\
                 &norm (10$^{-2}$)              &... &... &...   &1.5$^{+3}_{-0.1}$ \\
\\
\textsc{Cflux}   &F$_{\rm refl}^b$            &...   &0.42$\pm$0.01          &0.32$\pm$0.01     &0.43$\pm$0.02     \\
                &F$_{\rm total}^b$          &1.85$\pm$0.01        &1.90$\pm$0.01          &1.88$\pm$0.01     &1.89$\pm$0.01     \\
      &$\chi^2_{\nu}$ (dof)          &1.14 (2556)       &1.23 (2564)  &1.21 (2563)     &1.19 (2562)      \\
\hline
\end{tabular}
\\
\flushleft
Note: Uncertainties are reported at the 90\% confidence interval and 
were computed using MCMC (Markov Chain Monte Carlo) of length 
100,000. $^a$Equivalent hydrogen column density in 10$^{22}$ cm$^{-2}$; 
$^b$The 1--80 keV unabsorbed flux in 10$^{-7}$  ergs cm$^{-2}$ s$^{-1}$.
The folding energy E$_{\rm fold}$ is from the high energy cutoff power-law (HECUT) 
model, while F$_{\rm total}$ and F$_{\rm refl}$ represent the measured total source flux and  
the flux from the reflection component, respectively. Here, $\sigma_{\rm 6 keV}$ stands 
for	Gaussian width at 6 keV.
\\
}
\label{spec-para}
\end{table*}
\end{singlespace}


Typically, a reflection spectrum, originating from optically
thick material or the accretion disk, peaks at $\sim$20--30 keV
in the form of a prominent broad feature, known as the Compton hump. 
If the reflection arises from a medium close to the compact object, 
the strong gravitational field broadens the Fe-line feature and produces 
a skewed profile \citep{Fabian2000}, as seen in the present study. 
 To account for possible disk reflection in an X-ray pulsar system, 
we applied the \textsc{Coplrefl} model \citep{Ballantyne2012} on the absorbed 
high energy cutoff power-law (\textsc{HECUT}) continuum with a blackbody component
in our study. The \textsc{Coplrefl} code generates a reflection spectrum from a 
constant density disk (10$^{19}$ cm$^{-3}$) illuminated by a hard power-law with a 
variable high energy cutoff. This model combination is able 
to explain the hard X-ray spectral curvature as well as the observed iron 
line moderately well. The fit improves further by adding a \textsc{gsmooth} 
component subject to Doppler broadening to the \textsc{Coplrefl} code. 
We then allow the gravitational redshift to vary.  The goodness of fit 
is obtained as $\chi^2$/$\nu$ = 1.23 for 2578 dof with this model 
(Model-II). We note that the reflection component carries a significant 
part, approximately 20\%, of the source emission. Such a high reflection fraction indicates a 
larger covering area of reflecting  material in close proximity to 
the X-ray pulsar.

To investigate further, physically motivated new reflection models 
such as \textsc{xillver} \citep{Garcia2013} and \textsc{relxill} 
\citep{Garcia2014} were also combined with Cutoff+BB model. 
These codes assume emission from the inner accretion disk, and 
are widely used to explore the iron line and reflection components 
from  black holes and weakly-magnetized neutron stars in low mass 
X-ray binaries.   We first applied the angle-dependent reflection 
code, \textsc{xillver}  with Cutoff+BB model. As the above reflection 
code assumes a power-law model with a high energy cutoff, the cutoff energy 
was tied  to the  cutoff power-law continuum model.  This yields a fit 
with $\chi^2$/$\nu$ = 1.29 for 2564 dof. We then convolved with the \textsc{gsmooth} 
component as follows: {\tt tbabs*(gsmooth*xillver+cutoff+bbodyrad)} 
(Model-III). This improves the fit with a reduced-$\chi^2$ of 1.21 for 2563 dof.
Model-III also rectified the excess residuals at the ends 
of the 4--8 keV band as seen in the second panel of Figure~\ref{spec}, 
which also shows how the complex shape of the broad iron line and reflection 
components affect the energy continuum.

A more self-consistent inner-disk reflection code, \textsc{relxill} \citep{Garcia2014},
is used together with the Cutoff+BB base model 
(Model-IV). The purpose of this model is to test the role of reflection 
and blurring components in the Fe band. The \textsc{relxill} model is 
the currently most advanced version of reflection code that incorporates 
\textsc{xillver} and a relativistic blurring code \textsc{relline} 
together.  The model components of \textsc{relxill} are: emissivity 
index ($q$) for the disk, dimensionless spin parameter $a = cJ/GM^2$, 
inclination angle $i$ (in degrees), the ionization parameter log($\xi$), 
iron abundance with respect to the solar value $A_{Fe}$, and reflection 
fraction $f_{refl}$, inner disk radius  $R_{in}$ in units of gravitational 
radius $R_g$ ($GM/c^2$), outer disk radius $R_{out}$ in units of $R_g$, 
redshift $z$, and normalization. We allowed the inner disk radius $R_{in}$ 
to vary by fixing the outer radius at 990~$R_g$. The emissivity index for 
the disk was kept at 3 assuming isotropic point emission (see, e.g., 
\citealt{Wilkins2018}). The spin parameter is likely to be  
$<$0.7 for the range of neutron star masses and radii \citep{Miller2011_spin}, 
and is considered to be zero in our analysis \citep{Ludlam2018}. The choice of 
spin value does not affect our results much. We obtained only a difference of 
2~$R_g$ in the inner disk radius for the parameter $a=0$ and $a=0.7$. The above 
model approximated the energy continuum well with a goodness of fit 
$\chi^2$/$\nu$ = 1.19 for 2562 dof (fifth panel of Figure~\ref{spec}). 

 In this fit the derived inner disk radius 
is $\approx$130--490~$R_g$ from the neutron star. We also detected the gravitational 
red-shift of 0.022 from the reflecting gas, allowing a significant Doppler 
broadening and redshifted Fe-line profile by $\approx$6600~km~s$^{-1}$. 
In general the \textsc{xillver} and \textsc{relxill} models consider 
radiation dominated by a classical accretion disk, which may not be 
realized here. Still, the successful application to the 1.2--79 keV 
spectrum of \source  indicates  strong reflection and gravitational effects 
in the iron line band. It is important to mention that the reflection 
fraction  between Model-II, Model-III, and Model-IV is almost the same. It is 
only possible when the optically thick region is located close to the pulsar 
and covers substantial central emission. 

Best fitting spectral parameters for Model-I, Model-II, Model-III, and Model-IV
are given in Table~\ref{spec-para}. We also searched for absorption 
like features that could originate from cyclotron resonance scattering 
\citep{Jaisawal2017, Staubert2019A} in the 1.2--79 keV band. The non-detection 
of any signature of a cyclotron line in the broad-band spectrum could be 
an indication that the magnetic field of the pulsar is either 
$>$8$\times$10$^{12}$~G or $<$10$^{11}$~G. But at least a low magnetic 
field is in tension with the magnetic field estimates derived from other 
methods, as explained in Section~\ref{sec:intro}. Note also that only 
$\sim$10\% of all known binary X-ray pulsars have an identified cyclotron 
line feature \citep{Staubert2019A}, indicating that additional conditions 
besides a suitable magnetic field strength are required to form an 
observable line feature.

\section{Discussion}

We have analyzed \nicer observations of the bright transient 
pulsar  during its strong X-ray outburst in late 2017. 
The neutron star was accreting an order of magnitude above the 
expected Eddington limit at the outburst peak. The detection of the 
source at such a high luminosity has established it as the first 
Galactic ULX pulsar \citep{Tsygankov2018, Wilson2018}. Therefore, 
the study of \source\ provides an opportunity to explore the connection between 
accretion-powered X-ray pulsars and pulsating ULXs, and also to test 
theories of super-Eddington accretion.

The present paper reports the detection of iron emission lines 
in the 6--7 keV range and their evolution with luminosity 
during the giant outburst of \source.  A narrow 6.4~keV 
line is observed during the sub-Eddington state, as usually 
seen in other high-mass X-ray binary pulsars (see, e.g., 
\citealt{Jaisawal2016, Jaisawal2016smc}).  
The line-emitting region, in this phase, is probably located relatively far 
from the neutron star, i.e., in the accretion disk, inside the magnetosphere, 
or the stellar wind from the companion star. As the luminosity of the pulsar 
approaches the Eddington limit (L$_{\rm Edd}$ 
$\approx$1.8$\times$10$^{38}$~erg~s$^{-1}$ 
for a 1.4 \msol neutron star), the line profile broadens with 
significant contributions from 6.67 keV (Fe XXV), and 6.97~keV 
(Fe XXVI) features. A highly asymmetric line profile is observed  
 at luminosities (0.7--1.2)$\times$10$^{39}$~erg~s$^{-1}$ in 
the 1--10 keV band of \source (Figure~\ref{fe-ratio}). The $\approx$6.67~keV 
emission is found to lead the Fe~K band in the ultraluminous state. 
Weakly magnetized neutron stars or black-holes in X-ray binaries, or 
supermassive black holes in active galactic nuclei, usually show relativistic 
line profiles from their inner accretion disks  in the vicinity of the
compact object (see, e.g., \citealt{Fabian2000, Miller2007, 
Bhattacharyya2010}). The detection of such a broad, skewed line from a 
magnetized pulsar indicates possible Doppler effects and gravitational 
redshift at the emission site  in the present case.

\begin{figure}
\centering
\includegraphics[height=2.95in, width=3.32in]{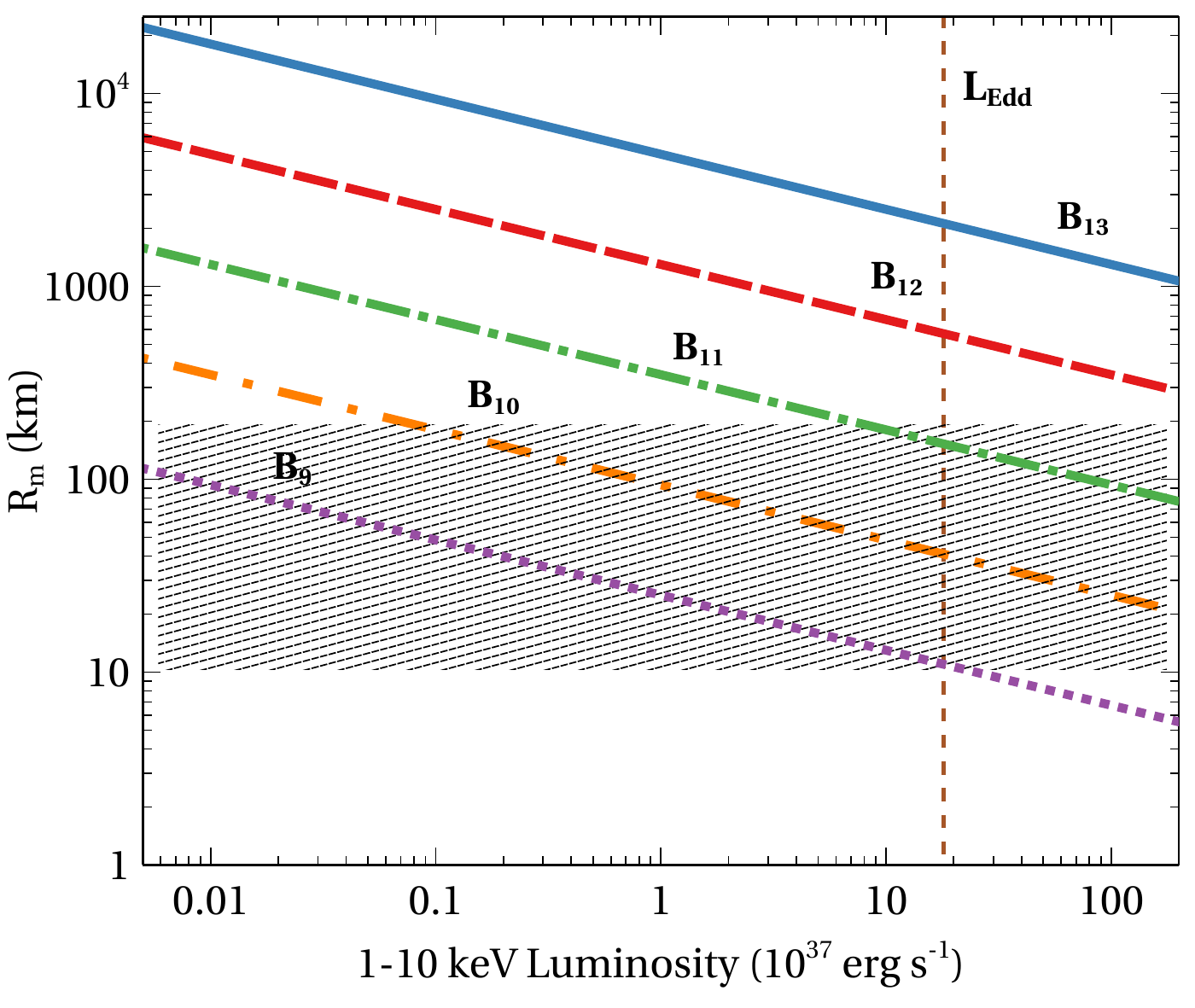}
\caption{Evolution of the magnetospheric radius (R$_{\rm m}$) 
with luminosity for a range of dipole magnetic 
fields. Here, B$_{\rm n}$ denotes the  field strength in 
units of 10$^{\rm n}$~G. The shaded area presents a zone up to a distance 
of $\approx$200 km from a canonical neutron star, allowing  
a relativistic iron line profile from the emitting medium 
due to gravitational effects. We observed a broad Fe-line 
clearly from \source in the super-Eddington regime. If the line 
originates from the  disk, a possible solution of the pulsar B-field 
can be obtained as an order of 10$^{11}$~G.} 
\label{inner-disk}
\end{figure}

To investigate the location of the emission sites, the magnetospheric 
radius, or inner accretion disk radius of the pulsar, is defined as 
R$_{\rm m}=$ 1300 L$_{37}^{-2/7}$~$M^{1/7}$~$R_6^{10/7}$B$_{12}^{4/7}$~km
for disk-accretion \citep{Ghosh1979, Mushtukov2017}, where $M$ is the 
mass of neutron star in  1.4\msol units, $R_6$ is the radius in units 
of 10$^6$~cm, and B$_{12}$ is the field strength in 10$^{12}$~G units. During 
the process of accretion, the magnetic field truncates the geometrically thick 
accretion disk near this point and channels the accreting material onto 
the neutron star poles. Using the above relation a large magnetosphere is 
expected at lower luminosities $\leq$10$^{37}$~erg~s$^{-1}$. It could be 
$\approx$5000~km for a highly magnetized pulsar 10$^{13}$~G (see 
Figure~\ref{inner-disk}). The iron line  is possibly produced 
by the accretion disk, the stellar wind, or from  optically thin plasma 
inside the magnetosphere in the sub-Eddington state.

 If we assume a  B field upper limit of 10$^{13}$~G and adopt
10$^{39}$~erg~s$^{-1}$ as the luminosity of \source, the implied magnetospheric  
is $\approx$1300~km from the central object (Figure~\ref{inner-disk}).  
The accretion disk would not produce a broad relativistic line in this situation. 
It is possible that the inferred strong magnetic field is dominated 
by a quadrupolar moment (see, e.g., \citealt{Bonazzola2015, Israel2017a}, 
and references therein), while the dipole  is weaker, allowing the 
disk to come closer to the  star. A line emitting region  
within 200~km likely carries a relativistic signature from the compact object. 
The above limit is obtained for a canonical neutron star with spin parameter $a=0$ 
by assuming neutral disk-reflection (see, e.g., \citealt{Bardeen1972, Garcia2014}), 
and is shown as a shaded area in Figure~\ref{inner-disk}. We note that 
a lower dipole B-field $\le$10$^{10}$~G would produce a broad line 
even at sub- and super-Eddington luminosities, while a more strongly magnetized 
neutron star, $B\ge10^{12}$~G, truncates the disk beyond 500 km at the 
Eddington limit. If we assume the observed broad Fe-feature comes from the 
disk, the dipole B-field of the pulsar should be of order 10$^{11}$~G 
(Figure~\ref{inner-disk}).

 The nature of the broad Fe-line from \source in the ultraluminous state is  
also investigated by using reflection codes, \textsc{xillver} and \textsc{relxill}. 
Our modelling suggests that the lines are emitted from a region close to the pulsar, 
and are subjected
to Doppler broadening (0.2~keV) and  redshift ($z$=0.022) which correspond 
to a net outflow velocity of $\approx$12,000~km~s$^{-1}$. This is in good agreement 
with the orbital velocity of the magnetosphere at 1300 km. We note that Model-II predicts
a much higher velocity of $\approx$20,000~km~s$^{-1}$, which is only attainable at 
a distance of $\approx$450~km from the X-ray source. Given this rapid flow, the bulk motion 
of material possibly combined with the gravitational effect near the neutron star gives 
rise to the broad Fe-line. Moreover, it is  possible that the receding inner parts 
of the disk are being illuminated more strongly than the approaching side. This may explain 
the observed asymmetric line in the super-Eddington state.

Alternative scenarios to describe the pulsar's observed spectral characteristics
are possible. Apart from the disk, another possibility of reflecting material
around magnetized ULXs is suggested  by \citet{Mushtukov2017}. The authors find that 
the optically-thin plasma trapped inside the magnetosphere becomes an optically 
thick curtain (envelope) in the ultraluminous stage $\sim$10$^{39}$~erg~s$^{-1}$. The photospheric 
emission from the envelope follows a quasi-thermal spectral shape with temperatures exceeding  
1~keV. This envelope is also large enough to reprocess the primary column emission from the 
neutron star,  similar to reprocessing in the disk. In extreme cases, the accretion curtains become thick enough 
to hide the pulsations, and apparently the ULX spectrum is described by a multicolor blackbody 
function without the column emission \citep{Mushtukov2017}. The presence of optically thick 
regions may also lead to the disappearance of cyclotron scattering features from the spectrum 
\citep{Mushtukov2017}. 

Still, the plasma trapped in the  magnetosphere remains 
optically thin closer to the accretion column, producing the emission line 
features \citep{Koliopanos2018}. The observed broad Fe-line from \source can likely originate 
from such accreted plasma located next to the star. Nevertheless, 
theoretical studies of super-Eddington accretion onto black holes or millisecond 
accreting pulsars also predicts massive ejections of the disk material either in 
the form of optically thick spherical outflows or optically thin ultrafast winds 
near the  axis of the compact object \citep{Poutanen2007, Parfrey2016, Sadowski2016}. 
These optically thin fast-moving outflows in close proximity to the neutron star 
are also potential sites for broad lines at high accretion phases.

It is important to note that this pulsar  launched a jet 
when the X-ray luminosity was $\ge$4$\times$10$^{36}$ erg~s$^{-1}$ 
during giant and fainter outbursts (for a distance of 7~kpc; 
\citealt{Eijnden2018, Eijnden2019}). At the same time, the luminosity-dependent 
emission lines were observed in the spectral range of \nicer. It is reasonable to consider 
that the emission lines could  originate in a jet because broad Doppler-shifted 
iron lines have previously been detected from a relativistic 0.26 c outflow 
in SS~433  \citep{Fabrika2015, Kaaret2017}. The optically thin medium within the
jet would also provide the Doppler shifted emission line, though, a relativistic 
line profile would  appear only if the jet or emission site is located  
next to the pulsar.

\section{Conclusions}

In this work, we have studied \nicer observations of \source during its 
2017--2018 outburst. The 1.2--10~keV energy spectrum from \nicer can be described 
by a cutoff power-law model plus Gaussian components for emission 
lines.  We have observed a narrow to broad line profile from
the pulsar during the sub- to super-Eddington states. Depending on the 
luminosity, the Fe band contained up to three emission lines 
of energies $\approx$6.4, 6.7, and 6.98~keV from neutral and highly ionized Fe-atoms. 
We have also shown that the iron line is relatively complex in the 
ultraluminous state,  affecting the continuum spectrum and the choice of 
spectral modeling. \citet{Tao2019} have described the  3--79 keV energy spectrum 
in the super-Eddington state with a cutoff power-law model with three blackbody 
components. With the inclusion, instead, of lines in the model and the soft X-ray coverage of \nicer, 
the 1.2--79 keV energy spectrum in this state is well described by 
a similar cutoff power-law plus blackbody model as used for the pulsar in the low luminosity 
state. We have shown that the two additional blackbody components 
used by \citet{Tao2019} are not required  at extreme luminosity as long as 
the iron-line complex is modeled carefully.  

Our analysis suggests that emission lines observed in sub-Eddington 
regimes mainly originate far from the pulsar. It could be hosted 
by the accretion disk, the plasma  trapped in the magnetosphere, 
optically-thin ejected material or in the jet outflow. We clearly observed  evolution 
of the line profile with luminosity in terms of the number of line 
constituents and their breadth, which suggests that the emission 
site was evolving across the outburst. In the beginning, these sites were 
located at a distance of $\sim$5000~km (for a luminosity of 10$^{37}$~erg~s$^{-1}$) 
and moved much closer to the neutron star during the ultraluminous 
phase. The detection of an asymmetric broad iron line at this stage 
also indicates the presence of Doppler effects and gravitational red-shift. 
The possibility of a broad line from the  disk can be considered  only if the neutron star
possesses a dipolar B-field strength in the narrow range between 10$^{11}$ 
and 10$^{12}$~G. The observed line in the super-Eddington state may also 
originate from an optically thin accretion curtain, an
ultra-fast plasma, or in the jet close to the compact object.

    
\acknowledgments
We sincerely thank the referee for useful suggestions on the paper. 
This work was supported by NASA through the \nicer mission and the
Astrophysics Explorers Program, and made use of data and software 
provided by the High Energy Astrophysics Science Archive Research Center (HEASARC).  
This project has received funding from the European Union's Horizon 2020 research 
and innovation programme under the Marie Sk{\l}odowska-Curie grant agreement No.\ 713683.
A.L.S. is supported by an NSF Astronomy and Astrophysics Postdoctoral Fellowship 
under award AST-1801792. D.A. acknowledges support from the Royal Society.


\facilities{ADS, HEASARC, \nicer, \nustar, \swift.}
\software{\textsc{HEAsoft} (v6.24), \textsc{XSPEC} (v12.10.0;  \citealt{Arnaud1996}), \textsc{Coplrefl}  \citep{Ballantyne2012}, 
\textsc{xillver} \citep{Garcia2013},  \textsc{relxill} \citep{Garcia2014},
Veusz.}


\bibliographystyle{fancyapj}
\bibliography{references}

\end{document}